# Red shift of spectral lines of hydrogen atoms caused by its electrical polarization

V. S. Severin

*Department of Theoretical and Applied Physics, National Aviation University,
1 Cosmonaut Komarov prospect, Kyiv 03058, Ukraine
E-mail: severinvs@ukr.net*

**Abstract.** A system of Lorentz oscillators is considered to interpret of spectral lines of hydrogen atoms. The dielectric permittivity of this system, which takes into account its electric polarization, is considered. The substance is examined in the gas state or in the form of small dust of matter. It is shown that the considering of the electric polarization results in a redshift of spectral lines of the substance and the appearance of dip of curve of spectrum. This dip takes place in the blue side with respect to the spectral position of the already shifted line. The magnitude of the red shift of spectral line and the width of this dip in the spectrum depend strongly on the concentration of hydrogen atoms that create this spectrum and on the spectral position of line that not shifted.

**Keywords:** spectral lines, optical properties, red shift.

PACS number(s): 32.70.Jz, 33.70.Jg, 51.70.+f, 98.62.Py

## 1. Introduction

The investigation of spectral lines of substance is the subject of optical spectroscopy since its origin, and it is used intensively in modern physics and astrophysics [1–5]. The Lorentz oscillator gives a simple but successful classical model of the spectral line and it is widely utilized to interpret spectral lines of substance [3–5].

Optical properties of substance are given by its dielectric permeability, expressed through the specific electric conductivity of this substance. However, the substance is characterized both by its external conductivity and by its internal conductivity. The dielectric permeability of substance is determined solely by the internal conductivity of this substance in accordance with the definition of dielectric permeability, and, therefore, optical properties of this substance are determined solely by the internal conductivity of this substance [6–11].

The external specific electric conductivity of substance is used traditionally instead of its internal specific electrical conductivity in the dielectric permittivity of substance. This substitution is a simplifying approximation that does not take into account the electric polarization of charge carriers of substance [12–17]. This approximation does not take into account the difference between the external electromagnetic field acting on the substance and the internal electromagnetic field acting in the substance.

Optical transitions of electrical charges in the substance are caused only by the internal electromagnetic field. The effect of the difference between external and internal electromagnetic fields is known in molecular spectroscopy. That is, the spectrum of individual molecules may differ significantly from their spectrum in condensed substance [1].

The paper is organized as follows: in section 2 optical properties of substance are modeled by the system of Lorentz oscillators taking into account the electric polarization of such a system. In section 3 obtained results are applied to the spectral lines of hydrogen atoms. General discussion of our results and conclusions are presented in section 4.

## 2. The influence of the electrical polarization of substance on its spectral lines

Optical spectra of vibrations of the crystal lattice of semiconductor solid solutions, depending on the composition of this solution, were considered in [16]. A system of identical Lorentz oscillators having the concentration $n$ was considered in [16]. It was shown in [16] that the dielectric permittivity $\varepsilon(\omega)$, which depends on the frequency of light $\omega$, of the system of Lorentz oscillators, in which the electric polarization of this system is taken into account, has the form

$$\varepsilon(\omega) = 1 + \omega_p^2 \frac{1}{\Omega_0^2 - \omega^2 + i\gamma\omega}. \tag{1}$$

Here the frequencies $\omega_p$ and $\Omega_0$ are given by the next formulas

$$\omega_p^2 = \frac{4\pi e^2 n}{m}, \tag{2}$$

$$\Omega_0^2 = \omega_0^2 \left(1 - \left(\frac{\omega_p}{\omega_0}\right)^2\right). \tag{3}$$

Where $m$ is the mass of oscillator and $e$ is its charge; $\gamma$ is the damping factor of oscillator; $\omega_0$ is the eigenfrequency of this oscillator before taking into account the electric polarization of the system of oscillators. The frequency $\Omega_0$ is the frequency of this oscillator after taking into account the electric polarization of the system of oscillators.

The dielectric permittivity of the system of Lorentz oscillators $\varepsilon_s(\omega)$, in which the electric polarization of this system is not taken into account, has the form [16]

$$\varepsilon_s(\omega) = 1 + \omega_p^2 \frac{1}{\omega_0^2 - \omega^2 + i\gamma\omega}. \tag{4}$$

The formula (3) gives the variation of the oscillator frequency $\Omega_0$ under the influence of the electric polarization of the system of oscillators. The frequency $\Omega_0$, which is given by the formula (3), decreases (has a red shift) when the concentration of oscillators $n$ increases. This theoretical result agrees with the experimental results for the optical spectra of vibrations of crystal lattice [16].

Let us consider the spectral line of an atom of substance. We believe that the dielectric permittivity of this substance is given by the formulas (1) – (4), in which $n$ is the atom concentration, $e$ is the electron charge and $m$ is the electron mass. In this case, $\omega_0$ is the frequency of spectral line of isolated atom (that is, the eigenfrequency of the oscillator without taking into account the electric polarization of the system of oscillators).

The real and imaginary parts of the dielectric permittivities $\varepsilon(\omega)$ and $\varepsilon_s(\omega)$ are given by the formulas

$$\varepsilon_1(\omega) \equiv \operatorname{Re}\varepsilon(\omega) = \frac{\left(\omega_0^2 - \omega^2\right)\left(\Omega_0^2 - \omega^2\right) + \gamma^2\omega^2}{\left(\Omega_0^2 - \omega^2\right)^2 + \gamma^2\omega^2}, \tag{5}$$

$$\varepsilon_2(\omega) \equiv -\operatorname{Im}\varepsilon(\omega) = \omega_p^2 \frac{\gamma\omega}{\left(\Omega_0^2 - \omega^2\right)^2 + \gamma^2\omega^2}, \tag{6}$$

$$\varepsilon_{s1}(\omega) \equiv \operatorname{Re}\varepsilon_s(\omega) = \frac{\left(\omega_0^2 - \omega^2\right)\left(\omega_0^2 - \omega^2 + \omega_p^2\right) + \gamma^2\omega^2}{\left(\omega_0^2 - \omega^2\right)^2 + \gamma^2\omega^2}, \tag{7}$$

$$\varepsilon_{s2}(\omega) \equiv -\operatorname{Im}\varepsilon_s(\omega) = \omega_p^2 \frac{\gamma\omega}{\left(\omega_0^2 - \omega^2\right)^2 + \gamma^2\omega^2}. \tag{8}$$

Light does not pass through substance at the frequency $\omega$, for which the real part of the dielectric permittivity is negative. Therefore, the optical spectrum of substance has a dip at this frequency. If there is the approximation $\gamma \to 0$, then the formula (5) gives that the condition

$\varepsilon_1(\omega) < 0$ is satisfied in the frequency interval

$$\Omega_0 = \sqrt{\omega_0^2 - \omega_p^2} < \omega < \omega_0. \tag{9}$$

On the other hand, the formula (7) gives that in the approximation $\gamma \to 0$, the condition $\varepsilon_{s1}(\omega) < 0$ is satisfied in the frequency interval

$$\omega_0 < \omega < \sqrt{\omega_0^2 + \omega_p^2}. \tag{10}$$

Consequently, the formulas (9) and (10) give that the model of the system of Lorentz oscillators, in which the electric polarization of this system is taken into account, and the model of the system of Lorentz oscillators, in which the electric polarization of this system is not taken into account, give the dip in optical spectrum in different frequency intervals.

If the frequency dependence of real part of dielectric permittivity is not taken into account, the place of spectral line in the spectrum is given by the maximum of imaginary part of dielectric permittivity. The formulas (6) and (8) give different locations of the spectral line in the spectrum, taking into account polarization and without it.

Let the wavelength of light of the spectral line, which corresponds to the Lorentz oscillator without taking into account the polarization, is $\lambda_0 = c2\pi/\omega_0$ ($c$ is the light velocity in vacuum); and the wavelength of light of the spectral line, which corresponds to the Lorentz oscillator taking into account the polarization, is $\lambda_1 = c2\pi/\Omega_0$. The spectral line of an isolated atom is located in the spectrum at the wavelength $\lambda_0$. Let the wavelength change is $\Delta\lambda = \lambda_1 - \lambda_0$.

The redshift parameter of spectral line $z(\lambda_0)$ is defined as $z(\lambda_0) = \Delta\lambda/\lambda_0$. The formula (3) gives

$$z(\lambda_0) = \frac{\lambda_1 - \lambda_0}{\lambda_0} = \frac{\omega_0}{\Omega_0} - 1 = \frac{1}{\sqrt{1 - \omega_p^2/\omega_0^2}} - 1. \tag{11}$$

Let $\lambda_p = c2\pi/\omega_p = \frac{c}{e}\sqrt{\frac{\pi m}{n}}$. The formula (11) gives

$$z(\lambda_0) = \frac{1}{\sqrt{1 - \lambda_0^2/\lambda_p^2}} - 1 = \frac{1}{\sqrt{1 - n/N(\lambda_0)}} - 1. \tag{12}$$

Here

$$N(\lambda_0) = \frac{\pi m c^2}{\lambda_0^2 e^2} = \frac{\pi}{r_0 \lambda_0^2} \tag{13}$$

is the characteristic concentration-dependent wavelength $\lambda_0$; $r_0 = e^2/(mc^2)$.

According to the formula (11), the line having the wavelength $\lambda = \lambda_0 + \Delta\lambda$ is observed in the spectrum of substance, if the condition $\omega_0 > \omega_p$ takes place. Otherwise, the limit $\omega_0 \to \omega_p$ gives $\Delta\lambda \to \infty$, and vibrations of oscillator disappear (i.e., the spectral line disappears). As a result, the blue boundary of observation of the spectral line of substance takes place. Namely, the observed wavelength $\lambda = \lambda_0 + \Delta\lambda$ of the spectral line of investigated substance is present in its spectrum only if the initial wavelength $\lambda_0$ is less than the wavelength $\lambda_p$:

$$\lambda_0 < \lambda_p = \sqrt{\frac{\pi}{r_0 n}} = \frac{c}{e}\sqrt{\frac{\pi m}{n}}. \tag{14}$$

The condition (14) gives the following restriction on the concentration of atoms $n$:

$$n < N(\lambda_0). \tag{15}$$

The condition (15) means that the observed wavelength $\lambda = \lambda_0 + \Delta\lambda$ of the spectral line of investigated substance exists in its spectrum only if the concentration $n$ is less than the concentration $N(\lambda_0)$, which depends on the wavelength $\lambda_0$.

## 3. The spectral line of hydrogen atoms

Let us consider the spectral line of hydrogen atoms from the point of view of the results presented in the previous section.

Generally, hydrogen exists in a laboratory in the form of molecules. On the other hand, hydrogen is present in an interstellar medium in the form of atoms having different concentrations. Moreover, there are numerous experimental results of their optical spectra.

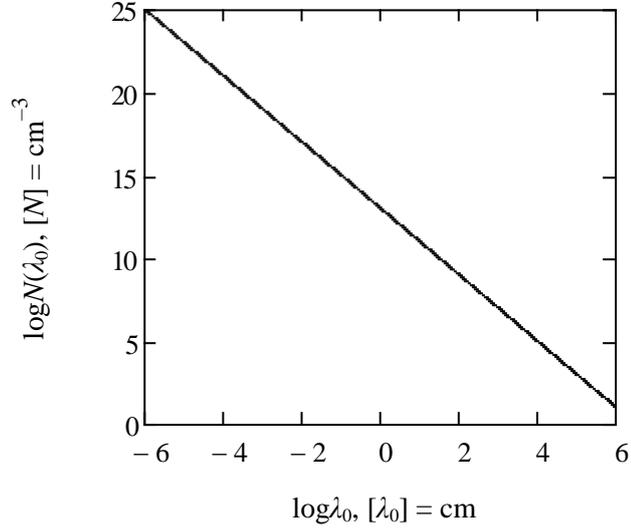

FIG. 1. The dependence of $N(\lambda_0)$ on the wavelength $\lambda_0$

As stated above, the wavelength $\lambda$, at which the spectral line of hydrogen atom is observed, depends essentially on the concentration of atoms $n$ and the characteristic concentration $N(\lambda_0)$, which is determined by the wavelength of spectral line of isolated atom $\lambda_0$.

Fig. 1 represents the dependence of the characteristic concentration $N(\lambda_0)$ on the wavelength $\lambda_0$. It can be seen that the value of concentration $N(\lambda_0)$ is very large for the optical range of wavelengths $\lambda_0$ and the corresponding value of the red shift of spectral line $z(\lambda_0)$ is very small for a rarefied atom gas.

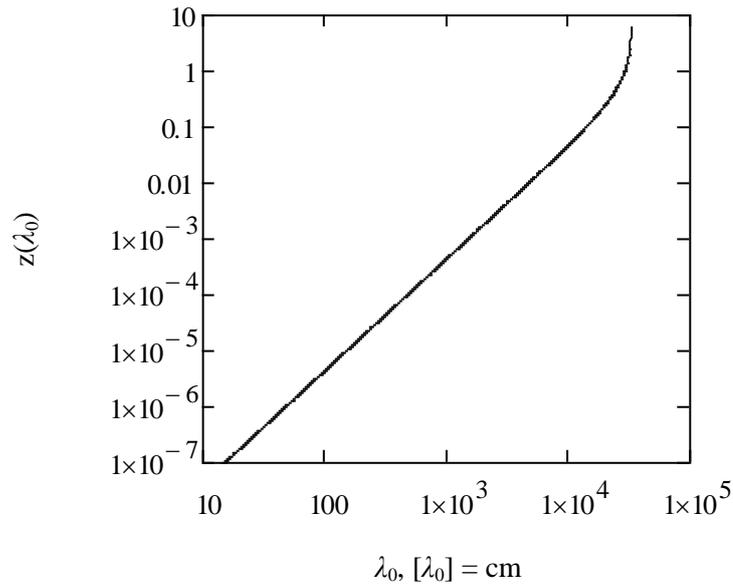

FIG. 2. The dependence of redshift parameter of spectral line $z(\lambda_0)$
on the wavelength $\lambda_0$ for the concentration value $n = 10^4$ cm$^{-3}$

The interstellar medium has the value of concentration of hydrogen atoms $n$ in the interval $10^{-1}$–$10^5$ cm$^{-3}$ (see Table 7.3 of [19]). For such a concentration, the red shift of spectral line may be significant in the radio-frequency range of wavelengths. For instance, the concentration $N(\lambda_0)$ is $1.12 \cdot 10^9$ cm$^{-3}$ for the wavelength $\lambda_0 = 10^2$ cm. The dependence of redshift parameter of spectral line $z(\lambda_0)$ on the wavelength $\lambda_0$ for the concentration value $n = 10^4$ cm$^{-3}$ is presented in Fig. 2. It is seen that the value of redshift parameter of spectral line $z(\lambda_0)$ strongly depends on the wavelength $\lambda_0$, and the concentration $n$.

The interstellar substance exists both in the form of gas and in the form of dust [2, 18-20]. Hydrogen is the main component of interstellar gas, stars and quasars [2, 18, 19]. Dust clouds are of great importance for the physical processes taking place in quasars [18].

Because of the low temperature of interstellar medium, hydrogen can be frozen on a dust surface [18-20]. The distinguishing feature of hydrogen atoms is their ability to penetrate into a solid with a large concentration, comparable with the concentration of atoms of solid, and, moreover, be in the atomic state [21]. The size of dust particles can vary from very small, corresponding to a cluster of a few tens of atoms, to fractions of a micron [18]. Depending on conditions, in which a dust particle is located, the concentration of hydrogen atoms $n$ in a dust particle can vary from a very small value to a value inherent to a solid substance.

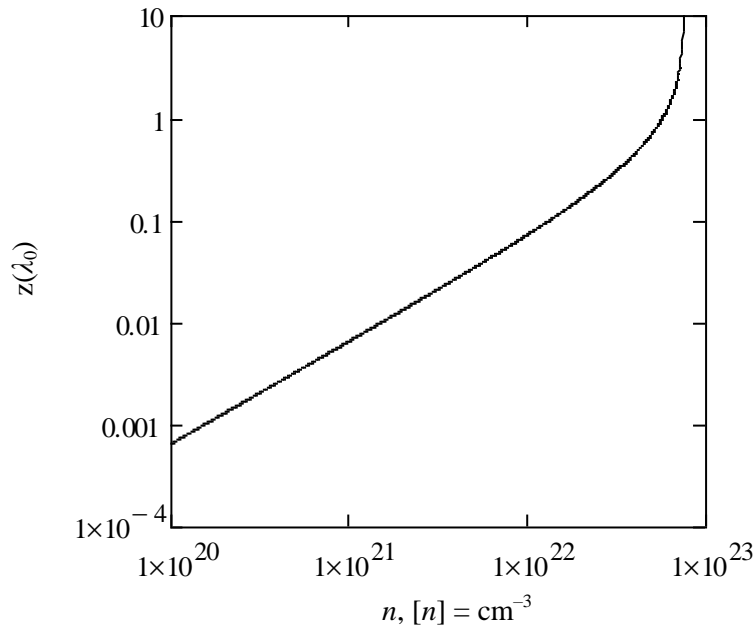

FIG. 3. The dependence of redshift parameter of spectral line $z(\lambda_0)$ on the value of the atom concentration $n$ for the wavelength $\lambda_0 = 1.216 \cdot 10^{-5}$ cm

The Ly$\alpha$ line of optical spectrum of an isolated hydrogen atom has the wavelength $\lambda_0 = 1.216 \cdot 10^{-5}$ cm. The formula (13) gives the value $N(\lambda_0) = 7.54 \cdot 10^{22}$ cm$^{-3}$ for this value of $\lambda_0$. Such a value of $N(\lambda_0)$ is close to the concentration of atoms in a solid. Solid bodies have its atom concentration of the order of $10^{23}$ cm$^{-3}$. Therefore, the formula (12) gives $z(\lambda_0) \gg 1$ for values of the concentration of hydrogen atoms $n$ in dust particles that are close to this value of $N(\lambda_0)$. The dependence of redshift parameter of spectral line $z(\lambda_0)$ on the value of the atom concentration $n$ for the wavelength $\lambda_0 = 1.216 \cdot 10^{-5}$ cm is presented in Fig. 3. As the concentration of hydrogen atoms $n$ in dust particles can reach large values, the dust cloud of quasar can give large values of the redshift parameter of spectral lines of hydrogen atoms.

## 4. Discussion and conclusions

If electric polarization of hydrogen atoms is taken into account, the red shift of spectral line of these atoms arises and the dip in their optical spectrum appears. This dip takes place in the blue side with respect to the spectral position of the already shifted line. The magnitude of the red shift of spectral line and the width of this dip in the spectrum depend strongly on the concentration of hydrogen atoms that create this spectrum and on the spectral position of line that not shifted.

This effect is significant in the radio-frequency range of wavelengths for interstellar hydrogen atoms.

In addition, the magnitude of this effect is considerable in the optical spectrum of hydrogen atoms, which are in clouds of interstellar dust. This is due to the possible large concentration of hydrogen atoms in the dust substance.

Quasars, for which dust clouds are of significant importance [22], give experimental spectra with such features [3, 22, 23]. For example, this takes place for the Ly$\alpha$ line of hydrogen atom (fig. 14.4 [3], fig.1 [22], fig. 8 [23]).

These results are of fundamental importance for astrophysics. Therefore, it is desirable to test them experimentally in a terrestrial laboratory. Indeed, under laboratory conditions, there are methods for obtaining and stabilizing atomic hydrogen, both in the free state and in a solid [21].